\documentclass[twocolumn,secnumarabic,amssymb, amsmath, nofootinbib,tightenlines,
nobibnotes, aps, prl]{revtex4}
\usepackage{amsmath}%
\usepackage{longtable}%
\usepackage{bm}%
\expandafter\ifx\csname package@font\endcsname\relax\else
 \expandafter\expandafter
 \expandafter\usepackage
 \expandafter\expandafter
 \expandafter{\csname package@font\endcsname}%
\fi

\usepackage{graphicx}
\usepackage{graphicx}

\begin{document}

\title{ Emergence of Multi-Scaling in a Random-Force Stirred Fluid. }
 \newcommand {\BU}{Department of Mechanical Engineering, Boston University, Boston, Massachusetts, 02215,USA}

\author{Victor Yakhot}
\affiliation{\BU}
\author{ Diego Donzis }
 \affiliation{Department of Aerospace Engineering, Texas A\&M University, College Station, Texas 77843,USA}



\date{\today}

\begin{abstract}
 \noindent We consider transition to strong  turbulence in an infinite fluid stirred  by a gaussian random force.  The   transition is {\bf defined}  as a   first appearance of anomalous scaling of  normalized    moments  of velocity derivatives (dissipation rates) 
 emerging from  the low-Reynolds-number   Gaussian background.  
 It is shown  that due to multi-scaling,  
 strongly intermittent  rare events     
  can be quantitatively described  in terms  of  an infinite number of different ``Reynolds numbers''  reflecting multitude of anomalous scaling exponents.   The theoretically predicted  transition disappears at   $R_{\lambda}\leq  3$.  The developed theory, is in a quantitative agreement with the outcome of large-scale numerical simulations.
\end{abstract}

\maketitle
PACS numbers 47.27\\

{\noindent \it  Introduction.}  
If an infinite  fluid is stirred by a gaussian random force supported in a narrow interval of the wave-numbers $k\approx 2\pi/L$, then a very weak forcing  leads to generation of  a random, close-to-gaussian,  velocity field.  In this flow the  mean velocity $\overline{\bf u}=0$ and one can introduce the large-scale Reynolds number $Re=u_{rms}L/\nu$ where the root-mean-square velocity  $u_{rms}=\sqrt{\overline{u^{2}}}$.
Increasing   the forcing  amplitude or decrease of viscosity result in a strongly  non-gaussian random flow with moments of velocity derivatives obeying the so-called anomalous scaling.  This means that the moments $\overline{(\partial_{x}u_{x})^{2n}}/(\overline{\partial_{x}u_{x})^{2}}^{n}\propto Re^{\rho_{2n}}$ where the exponents $\rho_{n}$ are,  on the first glance,  unrelated  ``strange'' numbers. 
In this paper we investigate  the  transition between these two different random/chaotic  flow regimes. First, we discuss some general aspects of the  traditional problem of hydrodynamic stability and transition to turbulence.\\
  
 \noindent Fluid flow can be described by the Navier-Stokes equations subject to boundary and initial conditions (the density is taken $\rho=1$ without loss of generality): 

\begin{equation}\partial_{t}{\bf u}+{\bf u\cdot\nabla u}=-\nabla p +\nu\nabla^{2}{\bf u} + {\bf f} \end{equation}

\noindent and   ${\bf \nabla\cdot u}=0$. The characteristic  velocity and length scales $u$ and $L$,  used for making  the Navier-Stokes equations  dimensionless, are somewhat arbitrary.  In the problem of  a flow past cylinder it is natural to choose $f=0$, $u=U$, and $L=D$ where $U$ and $D$ are free-stream velocity and cylinder diameter, respectively.  In a pipe/channel flow $u=U=\frac{1}{H}\int_{0}^{H}u(y)dy\propto u_{centerline}$ is the mean velocity averaged over  cross-section and $L=H$ is a half-width of the channel. 
In a fully turbulent flow in an infinite fluid one typically   takes 
$u=v_{rms}=\sqrt{\overline{v^{2}}}$ and $L$ equal to the integral scale of turbulence. Some other definitions will be discussed below.\\
Depending on  a setup,  a flow can be generated by pressure/temperature gradients,  gravity,  rotation, electro-magnetic fields etc represented  as forcing functions on the right side of  (1).
If viscosity $\nu\geq \nu_{tr}$ and the corresponding Reynolds number $Re=\frac{uL}{\nu}\leq Re_{tr}=\frac{uL}{\nu_{tr}}$, the solution to (1) driven by the regular (not random) forcing ${\bf f}$ is laminar and regular. As examples, we may recall parabolic velocity profile 
$u(y)$ in pipe/channel flows with prescribed pressure difference between  inlet and outlet. 
In this case the no-slip boundary conditions are responsible for generation of the rate-of-strain $S_{ij}=(\partial_{i}u_{j}+\partial_{j}u_{i})/2$. Another important example is  the so called Kolmogorov flow in an infinite fluid driven by the forcing function ${\bf f}=U(0,0,\cos kx)$.  In Benard convection the relevant regular patterns are rolls appearing as  a result of instability of  solution to the conductivity equation. 
Thus, the remarkably successful science of transition to turbulence deals mainly with various aspects of  non-equilibrium order-disorder  or laminar-to-turbulent transition. \\

\noindent    
In this paper we consider a completely different class of flows. In general,  the  unforced NS equations, being a very important and interesting object,  do not fully describe the  physical reality which includes   Brownian motion,  light scattering, random wall roughness, uncertain inlet conditions, stirring by ``random swimmers'' in biofluids   etc.   For example, a fluid  in thermodynamic equilibrium satisfies  the   fluctuation-dissipation theorem stating that there  exist an exact relation between viscosity $\nu$  in (1)  and a random noise ${\bf f} $ which is a Gaussian force  defined by the correlation function [1]:

 \begin{equation}\overline{f_{i}({\bf  k},\omega)f_{j}({\bf k'},\omega')}= (2\pi)^{d+1}D_{0}d(k)P_{ij}({\bf k})\delta(\omega+\omega')\delta({\bf k+k'})\end{equation}

\noindent  where the projection operator is: $P_{ij}({\bf k})=\delta_{ij}-\frac{k_{i}k_{j}}{k^{2}}$.   In  an equilibrium fluid    thermal fluctuations, responsible for Brownian motion   are generated by the forcing (2) with $D_{0}d(k)= \frac{k_{B}T \nu}{\rho}k^{2}\equiv D_{0}k^{2}$.  It is clear that, in general,  the function $d(k)$ in (2) depends on  the physics of  a  flow.

\noindent  The random-force-driven NS equation can be written in the Fourier space:
\begin{widetext}
\begin{equation}
 u_{l}({\bf k},\omega)=G^{0}f_{l}({\bf k,\omega})-\frac{i}{2}G^{0}{\cal P}_{lmn}\int u_{m}(q,\Omega)u_{n}(k-q,\omega-\Omega)d{\bf k}d\Omega
\end{equation}
\end{widetext}
\noindent where $G^{0}=(-i\omega+\nu k^{2})^{-1}$,  ${\cal P}_{lmn}({\bf k})=k_{n}P_{lm}({\bf k})+k_{m}P_{ln}({\bf k})$  and,  introducing the zero-order solution ${\bf u}_{0}=G^{0}{\bf f} \propto \sqrt{D_{0}}$, so that ${\bf u}=G^{0}{\bf f}+{\bf v}$,  one derives the equation for perturbation ${\bf v}$:

\begin{widetext}
\begin{eqnarray}
v_{l}(\hat{k})=-\frac{i}{2}G^{0}(\hat{k}){\cal P}_{lmn}({\bf k})\int v_{m}(\hat{q})v_{n}(\hat{k}-\hat{q})d\hat{q}\nonumber\\
-\frac{i}{2}G^{0}(\hat{k}){\cal P}_{lmn}({\bf k})\int  [v_{m}(\hat{q})G^{0}(\hat{k}-\hat{q})f_{n}(\hat{k}-\hat{q})+ G^{0}(\hat{q})f_{m}(\hat{q})v_{n}(\hat{k}-\hat{q})]d\hat{q} \nonumber \\
-\frac{i}{2}G^{0}(\hat{k}){\cal P}_{lmn}({\bf k})\int G^{0}(\hat{q}f_{m}(\hat{q})G^{0}(\hat{k}-\hat{q})f_{n}(\hat{k}-\hat{q})d\hat{q}
\end{eqnarray}
\end{widetext}
 
\noindent  where the 4-vector $\hat{k}=({\bf k},\omega)$.  If the equation (3)-(4) is driven  by a regular force or boundary and/or  initial conditions, then at low-Reynolds number  ($Re$)  it typically describes  a  regular (laminar) flow field ${\bf u}_{0}$  with  ${\bf v}=0$.  With increase of the  Reynolds number $Re\geq Re_{inst}$,  this zero-order solution can become unstable, meaning that initially-introduced small perturbations  $\bf v$ grow in time.  Further increase of Re leads  first to  weak interactions between the modes describing the ``gas'' of these perturbations and, eventually,  when $\frac{Re-Re_{inst}}{Re_{inst}}\gg 1$ mode coupling described by equation (4) becomes very strong.  This regime we call ``fully developed'' or strong turbulence. The problem of hydrodynamic stability is notoriously difficult and we know very little  about structure of solution for perturbations in the non-universal range $Re\approx  Re_{inst}$.  \\

\noindent Here we are interested in a simplified   problem of a flow generated by a  gaussian random force (2) with  a well -understood zero-order solution ${\bf u_{0}}=G^{0}{\bf f}$ which is {\bf not}  a result of an instability of a regular  laminar flow but  is prescribed by a choice  of a random force (2).  
The advantages of this formulation are clear from (4)  describing  the dynamics of perturbation 
${\bf v}$ driven by  an induced forcing given by the  $O(f^{2})\propto D_{0}$ last term in (4).  It is easy to see [1],[9] that dimensionless expansion parameter, related to a Reynolds number (see below),  is $\Gamma^{2}_{0}=\frac{D_{0}L^{4}}{\nu^{3}}\Delta$ where $\Delta=\int d(k)d{\bf k}$.  and, since we keep $L=O(1)$,  $\nu=O(1)$ and $\Delta=O(1)$, the variable  forcing amplitude $D_{0}$ can be treated as a dimensionless expansion parameter. 
Thus,  as $D_{0}\rightarrow 0$,  all  contributions to the right side of (4) can be neglected  and,  if   ${\bf f}$ stands for  the gaussian random function, then the lowest-order solution ${\bf u}_{0}$  is a gaussian field.. 
 However,   there always exist low-probability  rare events with $|{\bf v}|\geq |{\bf u}_{0}|$ responsible for the 
 strongly non-gaussian tails of the PDF.  Thus, in this flow gaussian velocity fluctuations coexist with the low-probability powerful events where substantial fraction of kinetic energy is dissipated. 
 At  even higher  Reynolds numbers (see below) the non-linearity in (4)  dominates the entire field.  This  complicated dynamics has been observed in experiments on a channel flow with rough (``noisy'') walls [2].\\
  
\noindent    {\it This regime is characterized by  the generation of velocity fluctuations ${\bf v}({\bf k,t})$ in the wave-number  range   $k>2\pi/L$  where the ``bare'' forcing ${\bf f}({\bf k})=0$,    which  is  the hallmark   of turbulence.      The above example shows that at least in some range of the Reynolds number    
 low and high-order moments may describe  very different  physical phenomena.  The transition between these two chaotic/random  states of a fluid  is a topic we are interested in this paper.  } \\

\noindent  Two  cases  are of  a special interest. 
 In  the  low Reynolds number regime  (below transition),  when $R_{\lambda}=\sqrt{\frac{5}{3{\cal E}\nu}}u_{rms}^{2}<R^{tr}_{\lambda}$,    the integral ($L$), dissipation ($\eta$)  and Taylor ($\lambda$) length scales are of the same order.    Therefore,   $(\partial_{x} u_{x})_{rms}=  (u(x+\eta)-u(x))_{rms}/\eta\approx (u(x+L)-u(x))_{rms}/L$  and, since we are interested in  instability of a gaussian flow,  the moments  
 
 $$M_{n}^{<}=\frac{{\overline{(\partial_{x}v_{x} )^{2n}}}}{\overline{(\partial_{x}v_{x})^{2}}^{n}}=(2n-1)!!$$

\noindent  independent on the Reynolds number.  In this case, since the $2n^{th}$-order   moment  can be expressed in powers of the variance,  this means that $(\partial_{x}v_{x})_{rms}$ is a single parameter  (derivative scale) representing statistical properties the flow   in this regime.
This is not always the case. The rms velocity derivative in high Reynolds number turbulent flows,   $(\partial_{x}v_{x})_{rms}=\sqrt{\overline{(\partial_{x}v_{x})^{2}}}$  is   {\it only one of an infinite number of  independent parameters needed to describe the field and in the vicinity of transition $Re\geq Re^{tr}$ }:

$$M^{>}_{n}= \frac{{\overline{(\partial_{x}v_{x} )^{2n}}}}{\overline{(\partial_{x}v_{x})^{2}}^{n}}= (2n-1)!!C_{n}Re^{\rho_{n}}\approx (2n-1)!!R_{\lambda}^{\rho_{n}}
$$

\noindent where $R_{\lambda}>R^{tr}_{\lambda}$ and  the proportionality coefficients  $C_{n}=O(1)$  [3], [4].  

\noindent {\it Below, this anomalous state of a  fluid we call strong turbulence as opposed to the close-to-gaussian low Reynolds number flow field, considered above.}  
In a  transitional,  low Reynolds number,  flow  we are interested in here, the forcing, Taylor and dissipation scales are of the same order $L\approx \eta\approx \lambda$. 
The   Reynolds number  based on the Taylor length-scale is thus:

\begin{equation}R_{\lambda}\equiv  R_{\lambda,1}= \sqrt{\frac{5}{ 3{\cal E}\nu}}v_{rms}^{2}\approx  \sqrt{\frac{5L^{4}}{3{\cal E}\nu}}\overline{({\partial_{x}v_{x})^{2}}}\end{equation}

\noindent The physical meaning of this parameter can be seen readily: multiply  and divide (5) by $\nu$ and by the dissipation scale $\eta^{2}$. This gives 

$$R_{\lambda}\propto  \frac{L^{2}}{\eta^{2}}
\times \eta^{2}\sqrt{\frac{\overline{{\cal E}}}{\nu^{3}}}\approx \frac{L^{2}}{\eta^{2}}$$

\noindent where $\eta^{4}\overline{{\cal E}}/\nu^{3}=O(1)$.  The effective Reynolds number $O(L^{2}/\eta^{2})$, which is the measure of  the spread   of the inertial range in $k$-space, is a coupling constant, familiar from dynamic renormalization group applications to randomly stirred fluids. 
To describe strong turbulence, one must introduce  an infinite number of   ``Reynolds'' numbers 
\begin{equation}
R_{\lambda,n}=  \sqrt{\frac{5L^{4}}{ 3{\cal E}\nu}} \overline{(\partial_{x}v_{x}))^{2n}}^{\frac{1}{n}}\propto R_{\lambda}^{\frac{\rho_{2n}}{n}}\propto \frac{L^{2}}{\eta^{2}}\frac{\overline{{\cal E}^{n}}^{\frac{1}{n}}}{\overline{{\cal E}}}
\end{equation}

\noindent where close to transition points  where $\eta\approx L$ we set $R_{\lambda}\equiv R_{\lambda,1}\approx Re$. The expressions for exponents $\rho_{2n}$ 

\begin{equation}\rho_{2n}=2n+\frac{\xi_{4n}}{\xi_{4n}-\xi_{4n+1}-1}; \hspace{2cm} \xi_{n}=\frac{0.383n}{1+\frac{n}{20}}\end{equation}

\noindent derived in the ``mean-field approximation''  in  [4]-[5], 
agree extremely  well with all available experimental and numerical data (see  Refs.[5]-[8]).   Theoretical predictions of anomalous exponents in a random-force-stirred fluid are compared with the results of numerical simulations  [6] on a top panel of Figure 1.  Note that normalized moments of dissipation rate $M_{n}({\cal E})$ are simply $M_{2n}$ in the present formulation. 
The same  exponents have been observed in a channel  flow [7] and Benard convection [8], indicating universality of small-scale features in turbulent flows.

\begin{figure}[h]
\includegraphics[height=8cm]{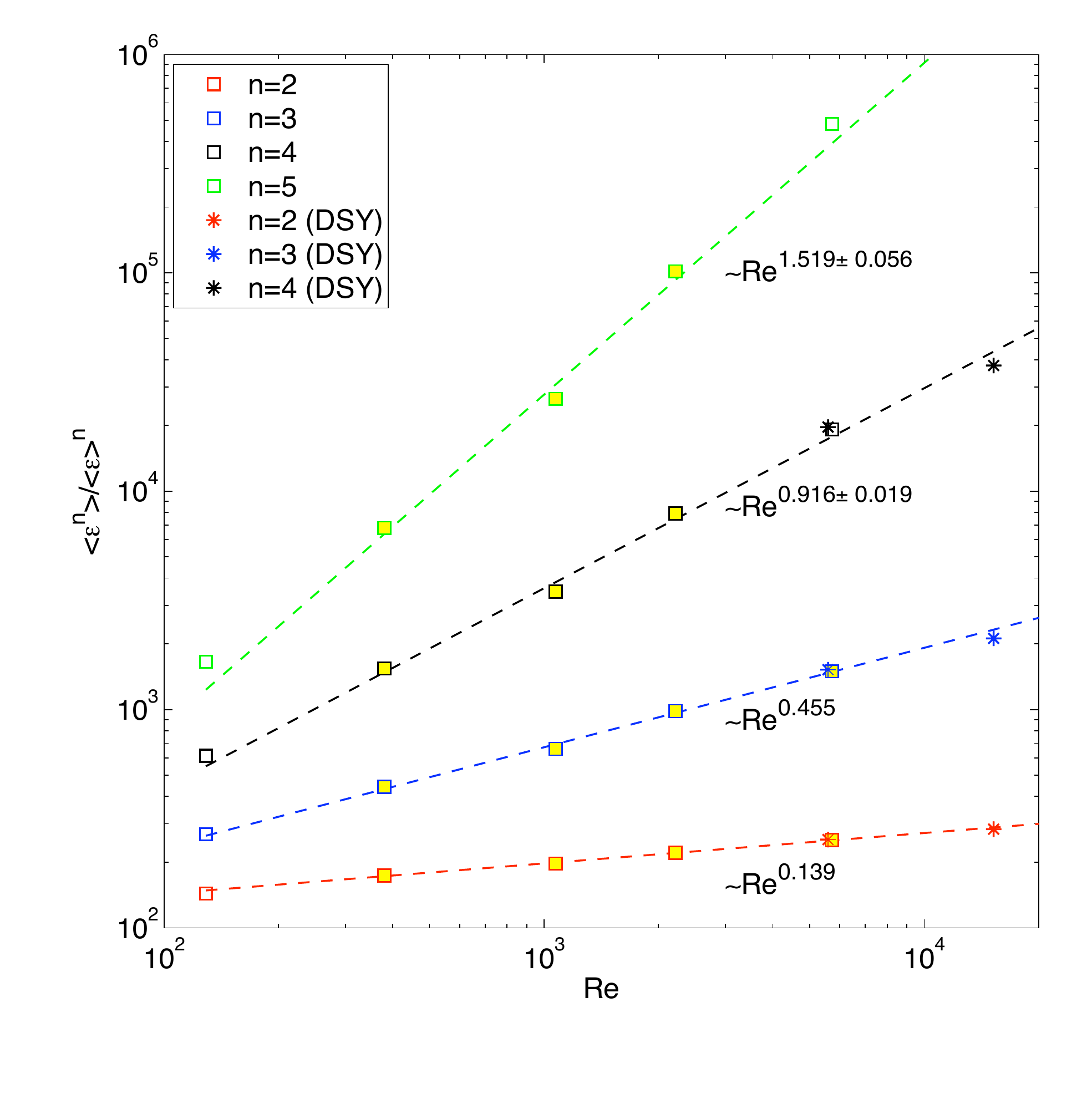}
\includegraphics[height=6cm]{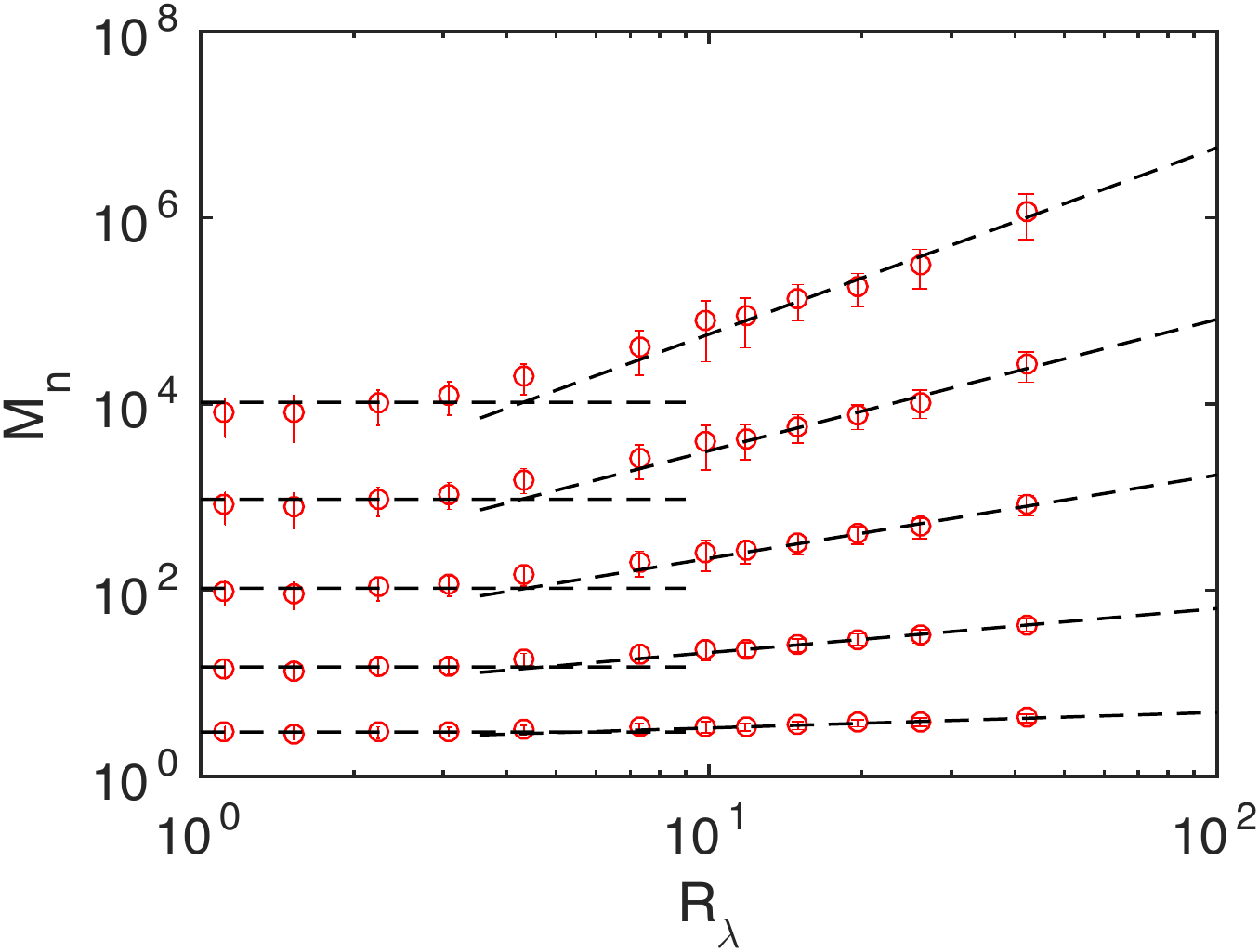}
\caption{ Top panel: Normalized moments $M^{{\cal E}}_{n}=\overline{{\cal E}^{n}}/\overline{{\cal E}}^{n}$ as a function
of Reynolds number.   Dashed lines: theoretical predictions and numerical simulations of Refs.[4]-[5];  Squares are from Ref.[5] and asterisks from our large DNS data base (see Ref. [6] );   Bottom panel:  Transition. 
The same moments in the  low-Re  transitional range  (present work)}
\end{figure}


\begin{figure}[h]
\includegraphics[width=8cm]{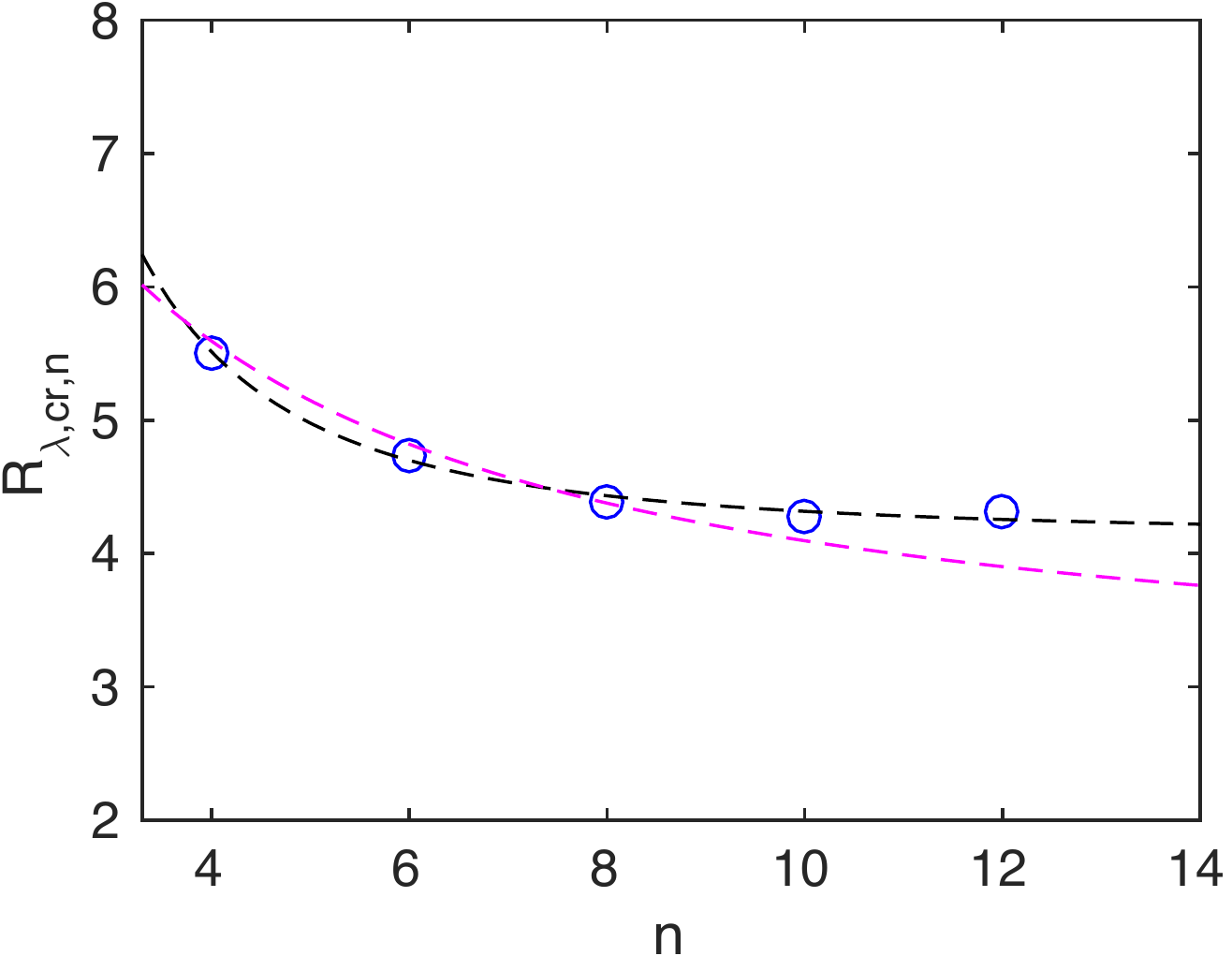}
\caption{ Transitional Reynolds number $R_{\lambda,1}(n)$  of the $n^{th}$ moment of velocity derivative.      Blue: numerical simulations of present work.  Red: Theoretical prediction with $R^{tr}_{\lambda,n.}=const=8.5$.}
\end{figure}
 
 {\it Transition between gaussian and  anomalous flows.}   
  In this paper  transition to turbulence is identified with   first appearance of non-gaussian anomalous  fluctuations of velocity derivatives.  
  The concept is illustrated on the bottom  panel of   Fig.1,  where  moments of velocity derivatives from well resolved numerical simulations (described below)  are plotted  against Reynolds numbers $R_{\lambda}\equiv R_{\lambda,1}\geq 2.$.  We can see that transition points of different moments, expressed in terms of $R_{\lambda}\equiv R_{\lambda,1}$,  are different and below we denote them $R^{tr}_{\lambda,1}(n)$.   
 It is important that transition point for  the lowest  order  moment $M_{n}$ with $n\approx 1$  has been found at $R_{\lambda}\equiv R_{\lambda,1}\approx 9$ first  discovered  in Ref.[5]   and analytically derived in [9]-[10].  This result can be explained as  follows.\\
  
\noindent In accord with the widely accepted methodology, consider the $R_{\lambda,1}\equiv R_{\lambda}$-dependence of  the normalized $n^{th}$  derivative moment  $M_{n}$ in a flow driven by   a relatively  weak force  $f$ and large viscosity $\nu$.  Then,   gradually decreasing viscosity,  one reaches the critical   magnitude $\nu=\nu_{tr}$  corresponding to $R_{\lambda}^{tr}(n)=R^{-}_{\lambda}(n)$  which is the upper limit  for gaussianity of the $n^{th}$ moment.
Then,  consider  the same flow but at a    very large Reynolds number  (small viscosity). In this,  strongly turbulent case,   the large-scale low-order moment,   $M_{4}$ for example,  are dominated by a  huge  turbulent viscosity $\nu_{T}\propto {\cal E}^{\frac{1}{3}}L^{\frac{4}{3}}$, the largest effective viscosity , accounting for velocity fluctuations at the scales $r<L$  [1].  
 The effective Reynolds number, corresponding to the integral scale $L$,   is   $R_{\lambda}^{+} \propto \sqrt{L^{4}/({\cal E}\nu_{T}(L))}(\partial_{x}u)_{rms}^{2}$.  
 This way one reaches the smallest possible Reynolds number $R_{\lambda}\approx 9$ of strongly turbulent (anomalous) flow (see Fig.1).
 If, in accord with experimental and numerical data, we assume   that transition is smooth and at a transition point the Reynolds number is a continuous function meaning that  $R_{\lambda}^{-}=R_{\lambda}^{+}$, where $R_{\lambda}^{\pm}$ stand  for the magnitudes just above and below transition, 
   we can write:
 
$$R_{\lambda}^{tr}(4)=\sqrt{\frac{5}{3{\cal E}\nu_{tr}}} v_{rms}^{2}=\sqrt{\frac{5}{3{\cal E}\nu_{T}(L)}} v_{rms}^{2}$$  

\noindent where effective viscosity  of turbulence at the largest (integral) scale calculated in 
 Refs. [9]-[11], is given by 
 
 \begin{equation}
\nu_{T}\equiv  \nu(L)\approx 0.084\frac{{\cal K}^{2}}{{\cal E}};   \end{equation}\\

 \noindent where ${\cal K}=v_{rms}^{2}/2$ stands for kinetic energy of velocity fluctuations.  Substituting this into the previous relation gives:

\begin{equation}R_{\lambda}^{tr}(4)=\sqrt{\frac{5}{3{\cal E}\nu}} v_{rms}^{2}=\sqrt{20/(3\times 0.084)} =8.98\approx 9.\end{equation}

\noindent  extremely close to the outcome of numerical simulations. 
The coefficient $C_{\mu}=0.084$, derived in [9]-[11] is to be compared with $C_{\mu}=0.09$ widely used in engineering turbulent modeling for  half a century [12].  It follows from the  relations (5)-(6):

\begin{equation}
R^{tr}_{\lambda}(n)\equiv R^{tr}_{\lambda,1}(n)=(R^{tr}_{\lambda,n})^{\frac{n}{\rho_{2n}}}
\end{equation}

\noindent   The Reynolds number dependence of normalized moments  of velocity derivative is shown on Fig.1.  The data in the bottom panel  of Fig.1 was generated from  a new set of simulations 
at very low Reynolds numbers.
As in [6],
numerical solutions to Navier-Stokes equations 
are obtained from Fourier pseudo-spectral
calculations with second-order Runge-€"Kutta integration in time. 
The turbulence is
forced numerically at the large scales,
using a combination of independent Ornstein-€"Uhlenbeck processes
with Gaussian statistics and finite-time correlation. 
Only low wavenumbers modes
within a sphere of radius $k_{F}\approx 2$
in wavenumber space are forced.
In order to obtain different Reynolds numbers, 
viscosity is changed accordingly while the forcing at 
large scales remains constant. 
In this approach, thus, large scales, and thus the energy
flux, remain statistically similar.
Resolution is at least $k_{max}\eta\approx 3$ at the 
highest Reynolds number which was found to produce converged 
results at the Reynolds numbers investigated here.

Velocity fields are saved at regular time intervals that are 
sufficiently far apart
(of the order of an eddy-turnover time) to 
ensure statistical independence between them. For each field
velocity gradients moments are computed and averaged over space. 
Ensemble average is computed across these snapshots in time
and are used to compute confidence intervals also shown in Fig.1. 
 
The intersection points of curves describing gaussian moments  
(horizontal dashed lines) and  those corresponding to the  fully- turbulent anomalous scaling 
  give transitional  $R_{\lambda}^{tr}(n)$  for each moment.  These are compared  to the theoretical prediction of Eq.(10) with $R_{\lambda,n}^{tr}\approx 8.5$ in Fig.2.
This result can be understood as follows:  in accord with theoretical predictions the  transitional  Reynolds number $R^{tr}_{\lambda,n}\approx const\approx 9$  in each statistical realization.
 If $R_{\lambda,1}<R^{tr}_{\lambda}\approx 9$,  the transition is triggered by the low -probability violent velocity fluctuations 
  $\overline{(\partial_{x}v_{x})^{n}}^{\frac{1}{n}}>(\partial_{x}x_{x})_{rms}$ coming    from the tails of probability density.   
   
  \noindent It is also interesting to evaluate the limiting, smallest,  transitional Reynolds number  following  (10)  in the limit $n\rightarrow \infty$. The relations (5)-(6),(10)  give $R_{\lambda}= R^{tr}_{\lambda,1}\rightarrow 2.92$.  Evaluated on a popular model $\xi_{n}= \frac{n}{9}+2(1-(\frac{2}{3})^{\frac{n}{3}})$ [13], one readily derives  $R^{tr}_{\lambda,1}\rightarrow 3.81$.  According to both models,  in a   flow with $R_{\lambda}\leq 3$,   no transition to strong turbulence defined by anomalous scaling of moments of velocity derivatives  exist. \\
  
  \noindent {\it Summary and conclusion.} In this paper a    problem of transition between two different {\it random } states has been studied both  analytically and numerically.  It has been shown that  while the gaussian state can be described in terms of the Reynolds number based on the  variance of probability density,  the description of the  intermittent state of {\it strong}  turbulence requires  an infinite number of "Reynolds numbers"  $R_{\lambda,n}$ reflecting  the multitude of anomalous scaling exponents of different-order moments  ($n$)  of velocity derivatives.  This   novel  concept   enables one to account  for   both typical  and violent extreme  events  responsible for emergence of anomalous scaling  in the ``sub-critical'' state when the widely used Reynolds number $R_{\lambda,1}<R^{tr}_{\lambda}$ is small.   It has also been demonstrated  that,  in accord with the theory,  the critical  $R^{tr}_{\lambda,n}\approx 9$ is   independent of $n$.  The proposed theory is in a good quantitative agreement with the  results of large-scale  direct numerical simulations presented above.   The role of turbulent bursts in  low Reynolds number flows in various physico-chemical processes  and the problem of  universality will be discussed in  future communications.

\begin{acknowledgements}     
We are grateful to  H. Chen,  A.Polyakov,  D. Ruelle, J. Schumacher,  I. Staroselsky, Ya.G. Sinai, K.R. Sreenivasan  and M.Vergassola    for many stimulating and informative discussions.  DD acknowledges support from NSF.
 \end{acknowledgements}         

\begin {references}
 \noindent 1. \  L.D.Landau \& E.M. Lifshits, ``Fluid Mechanics'', Pergamon, New York, 1982; \  D. Forster, D. Nelson \&  M.J. Stephen, Phys.Rev.A {\bf 16}, 732 (1977);\\
 \noindent 2. \ C.Lissandrello,  K.L.Ekinci \& V.Yakhot, J. Fluid Mech, {\bf 778}, R3 (2015);\\
\noindent 3. \ T. Gotoh \& T. Nakano , J. Stat Phys. {\bf 113},  855 (2003)  ; \\
\noindent   4. \ V.Yakhot, J.Fluid Mech.  {\bf 495}, 135 (2003);  \\
\noindent 5. \  J.Schumacher, K.R. Sreenivasan \& V. Yakhot, New J. of Phys.  {\bf 9}, 89 (2007); \\
\noindent 6. \  D.A. Donzis, P.K. Yeung \&  K.R. Sreenivasan,  
Phys.Fluids {\bf 20}, 045108 (2008); \\ 
\noindent  7. \   P.E. Hamlington, D. Krasnov, T. Boeck \&  J. Schumacher, 
J. Fluid. Mech. {\bf 701}, 419-429 (2012);\\ 
\noindent 8. \  J. Schumacher, J. D. Scheel, D. Krasnov, D. A. Donzis, V. Yakhot \&   K. R. Sreenivasan,
Proc. Natl. Acad. Sci. USA {\bf 111}, 10961-10965 (2014)\\
\noindent [9].  \ V. Yakhot \& L. Smith,  J. Sci.Comp. {\bf 7}, 35 (1992);\\
\noindent 10. \ V.Yakhot,, Phys.Rev.E, {\bf 90}, 043019 (2014);\\ 
\noindent 11 .  \ V. Yakhot, S.A. Orszag, T. Gatski, S. Thangam \& C.Speciale,   Phys. Fluids A{\bf 4}, 1510  (1992);\\
\noindent 12. \   B.E.   Launder,and D.B. Spalding. Mathematical Models of Turbulence, Academic
Press, New York (1972);  B.E. Launder and D.B. Spaulding,  
Computer Methods in Applied Mechanics and engineering, {\bf 3}, 269 (1974).\\
\noindent 13. \ Z.S.She \& E. Leveque, Phys.Rev.Lett. {\bf 72}, 336 (1994).
\end{references}

\end{document}